\documentclass{article}
\pdfoutput=1

\usepackage{graphicx}

\title{Study of the charge profile of thermally poled electrets}
\date{}
\author{
S.E. Parsa$^1$, J. Trull$^1$, X. Colom$^2$ and J. Sellar\`es$^1$\thanks{E--mail: {\tt jordi.sellares@upc.edu}} \\
$^1$Departament de F\'{\i}sica, $^2$Departament d'Enginyeria Qu\'{\i}mica, \\
Universitat Polit\`ecnica de Catalunya, \\
c. Colom~1, E-08222 Terrassa, Spain.
}

\begin{document}

\maketitle

\begin{abstract}
The charge profile of thermally poled electrets has been studied using two different methods, laser induced pressure pulse (LIPP) and pulsed electroacoustic (PEA), to gain insight into the mechanisms that are activated and assess which is the most appropriate method to study the charge profile. Disc--shaped PET samples have been conventionally poled to activate both the $\alpha$ and the $\rho$ relaxation and, right after, partially discharged up to a temperature $T_{\mathrm{pd}}$. In this way, samples with a different combination of dipolar and space charge polarization have been obtained. Both LIPP and PEA reveal asymmetric profiles for $T_{\mathrm{pd}}$ below the glass transition temperature, that progressively become antisymmetric for higher temperatures. The shape and evolution of the charge profiles can be explained assuming injection of negative carriers from the anode that enhances the trapping of positive carriers near this electrode. It can be observed that PEA is able to detect a wider variety of polarization mechanisms in the system while LIPP gives a simpler picture of the charge profile.
\end{abstract}

\section{Introduction}

An electret is made of a dielectric material that has been poled in such a way that it creates quasi--permanent external and internal electric fields \cite{eguchi25, gerhard87}. Electrets can be poled by several techniques such as corona charging, electron beam irradiation or thermal poling, among other ones \cite{gerhard87}. Thermal poling (TP) is a well-known technique to obtain electrets. It is the basis of techniques such as thermally stimulated depolarization currents (TSDC) \cite{chen81, sessler99}, that has been used for decades to study dipolar and space charge relaxation in solids. For this reason, the results of TP are well known from a macroscopic point of view. In spite of this fact, there is still unknown information about the microscopic mechanisms that are activated by TP. 

For example, it seems out of question that the $\alpha$ relaxation is the dielectric signature of the structural relaxation \cite{belana85} while being due to the reorientation of molecular dipoles and that the $\rho$ relaxation is due to free space charge trapped in localized states \cite{mudarra97}. These relaxations can be seen, for example, as peaks in a TSDC spectrum. Nevertheless, it is suspected that additional mechanisms are needed to fully explain these phenomena. A throughout understanding of the mechanisms that give rise to a relaxation is especially useful when interpreting data from dielectric spectroscopy and can also give clues on how to obtain more stable electrets or electrets that are better suited for particular applications.

One important piece of information related to the microscopic mechanisms is the charge profile of the electret \cite{montanari05}. Many studies have been made about the charge profile of corona charged or electron beam irradiated electrets but there are few of thermally poled electrets \cite{kazansky96}. This is because they tend to store less charge and be less stable than electrets poled by other means. As a consequence, its charge is more difficult to measure and results are less conclusive.

Several methods can be used to show the charge profile of an electret sample \cite{lewiner05} up to a certain resolution \cite{hole08}. Acoustic methods can give directly (without the need of deconvolution) the charge or field distribution in the sample. Additional quantities, such as surface potential, can be obtained by calculation from these ones \cite{chen07}. Depending on the signal--generation process, acoustic methods can be divided into LIPP (laser-induced pressure pulse), PPS (piezoelectrically-generated pressure step) or PEA (pulsed electro acoustic) methods. In all these cases, pressure waves generated  either at the sample surface or at charge layers in the bulk, propagate through the sample with the speed of sound waves. The deformations of the material that are produced cause currents or voltages at the electrodes due to charge displacement, changes in dielectric permittivity or the piezoelectric effect \cite{sessler97}.

Within the LIPP method \cite{sessler82}, one side of the sample is covered with a target and irradiated with a short laser pulse. The target absorbs the energy of the laser pulse and the rapid expansion that results produces a brief but intense pressure pulse in the dielectric. When the compressed region pulse travels through the sample a current is induced in the external circuit, due to non-uniform charge distribution and permittivity change \cite{gerhard83}, reflecting the charge distribution. We can record the charge profile connecting this setup with an oscilloscope. Allegedly, one of the advantages of the LIPP method with regards to other methods is that it produces pressure pulses with steeper rise fronts, which increases the spatial resolution of the system. The LIPP method has been used over the years extensively by their creators, Sessler and Gerhard, in their investigations of charge profile in thin-film polymer materials \cite{sessler86}. It can be used to determine space charge dynamics in charged samples \cite{sessler92, chong07}, for example, to assess the degradation of a dielectric \cite{mazzanti03}. 

Instead, the principle of the electroacoustic--pulse (PEA) method is based on the Coulomb law. An externally applied electric field pulse induces a perturbing force on the charges that are present in the material. This perturbation launches a sound wave which originates from the charge distribution. The acoustic signal is detected by a piezoelectric transducer placed on one of the electrodes \cite{maeno88}. The space-charge profile information contained in the signal is measured and calibrated through the use of digital signal processing \cite{chen06}. The main difference in contrast to other methods, such as LIPP where the pressure wave is generated externally, is that the acoustic wave is generated internally by the space charge \cite{ahmed97}. Initially deconvolution or other data processing was necessary to obtain the charge profile. Some very important improvements were made recently to the PEA method which eliminated the need for deconvolution \cite{li94}. The duration of the voltage pulse in PEA measurements is usually between 5 to 40~ns \cite{fleming05}.

The aim of this work is to study the charge profile of thermally poled electrets in a double way. On the one hand, we want to assess which method, LIPP or PEA, is more appropriate for this task, which are their relative merits and if the information that they provide is complementary in some way. On the other hand we want to know which are the mechanisms that are activated in the poling stage and how they are spatially distributed along the sample. To this end, we will measure the charge profile of samples prepared in such a way so that higher frequency mechanisms are relaxed leaving only progressively lower frequency mechanisms activated. We expect to find differences in the charge profiles that allows us to elucidate the role of dipolar (higher frequency) and space charge (lower frequency) mechanisms in TP and identify other relevant factors such as the role of the electrodes. 

\section{Experimental}

Experiments have been performed on amorphous PET samples. PET was supplied by Goodfellow in the form of sheets with a thickness of $230$~$\mu$m. 
Characterization by differential scanning calorimetry has shown that the glass transition takes place at $80$~$^\circ$C and crystallization begins around $100$~$^\circ$C. It has also been checked that the as received material has no noticeable crystallinity. 
Samples of $2.5 \times 2.5$~cm$^2$ were cut and aluminum electrodes with a diameter of $2$~cm were vacuum deposited on both sides.

\begin{figure}
\begin{center}
\includegraphics[width=8cm]{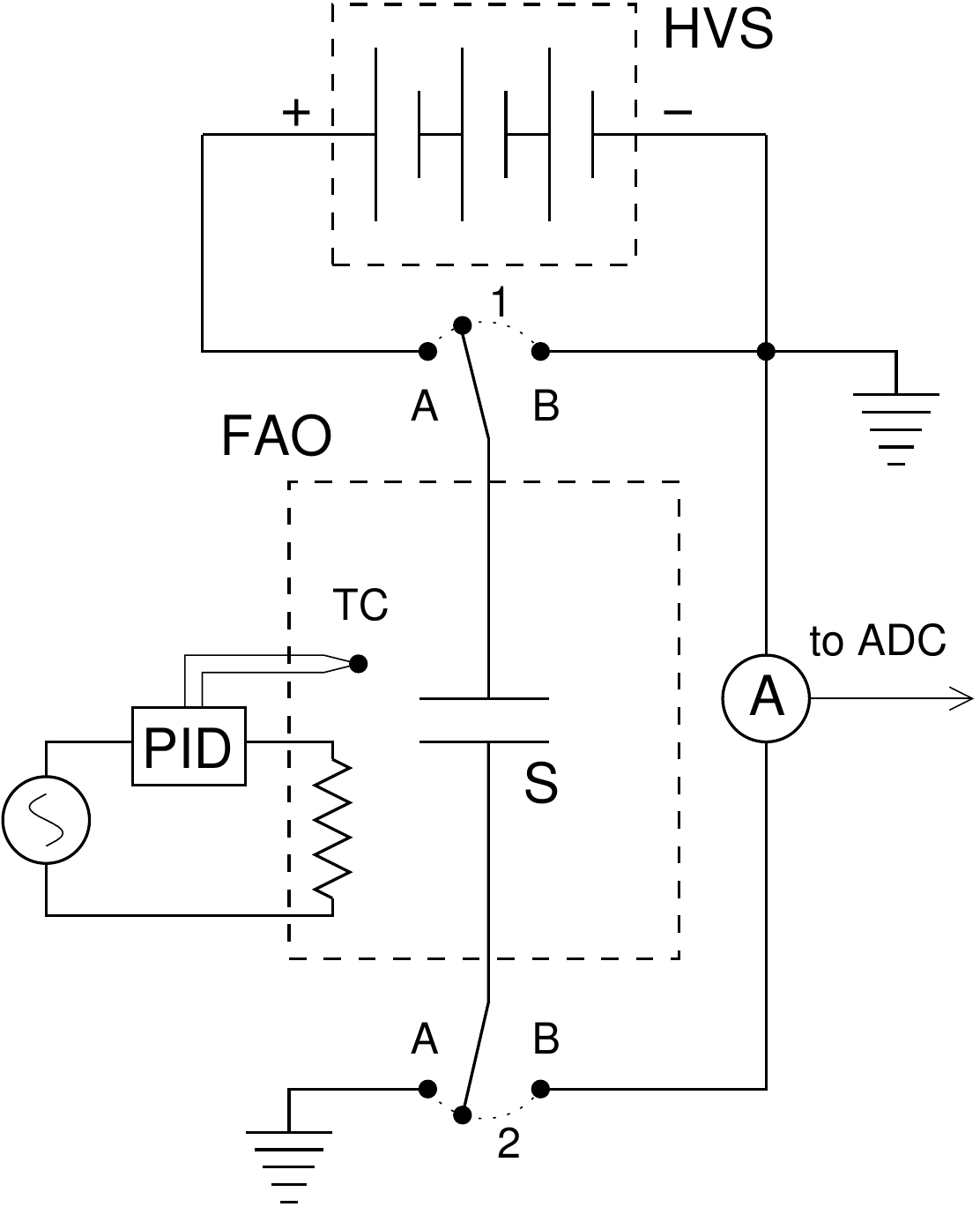}
\caption{Scheme of the TSDC setup.}\label{tsdc-setup}
\end{center}
\end{figure}

Prior to experiments, samples were thermally poled and partially discharged using a TSDC setup. An scheme of a TSDC setup can be seen in Figure~\ref{tsdc-setup}.
The sample (S) is poled by the high voltage source (HVS) setting the two--way switches~1 and~2 in position~A. Instead, switches should be in position~B to depole the sample and record the displacement current through the amperimeter. The sample is inside a forced air oven (FAO) driven by a PID controller with a thermocouple (TC) as input.  

\begin{figure}
\begin{center}
\includegraphics[width=12cm]{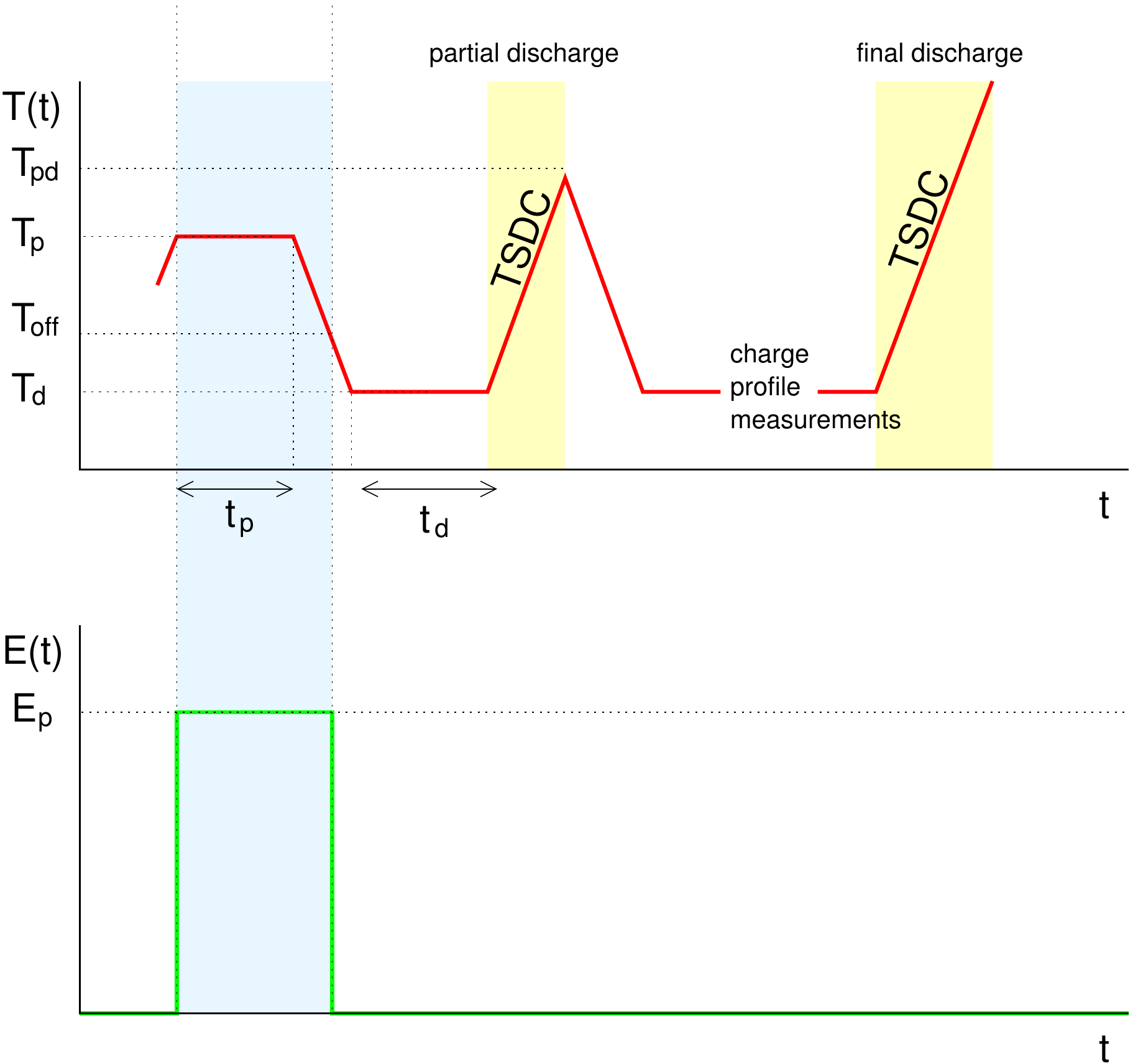}
\caption{Thermal and Electrical history of the sample in the experiments that have been performed.}\label{thermal-poling}
\end{center}
\end{figure}

The thermal and electrical history of our samples is described in Figure~\ref{thermal-poling}.
In first place, the sample is heated up to $95$~$^\circ$C, a temperature well-above the glass transition but not high enough to crystallize the sample significantly. In this way, the thermal history of the sample is erased and the experiment begins in a structural equilibrium state. To pole mechanisms within a broad range of frequencies, the sample is poled by applying a voltage of $1500$~V isothermally at a poling temperature ($T_\mathrm{p}$) of $95$~$^\circ$C for $5$~min ($t_\mathrm{p}$ in Figure~\ref{thermal-poling}) and, next, the sample is cooled at a constant rate of $2$~$^\circ$C/min down to a deposit temperature ($T_\mathrm{d}$) of $25$~$^\circ$C. The electric field is switched off during this cooling ramp, at $50$~$^\circ$C. This allows us to pole mechanisms with different relaxation times. In particular, we will pole the $\alpha$ relaxation, which is a dipolar mechanism, and the $\rho$ relaxation, which is a space--charge mechanism.

Since we want to compare samples where different mechanisms have been poled, we discharge the samples partially. After $15$ minutes ($t_\mathrm{d}$) at a deposit temperature ($T_\mathrm{d}$) of $15$~$^\circ$C the sample is heated up to a partial discharge temperature ($T_{\mathrm{pd}}$) at the same rate of $2$~$^\circ$C/min. Once this temperature is reached, the sample is cooled down to room temperature, also at $2$~$^\circ$C/min.

In this way, mechanisms that relax in a short time at temperatures below $T_{\mathrm{pd}}$ are depoled and do not contribute to the polarization of the sample. The only mechanisms that remain activated are the ones that remain poled at $T_{\mathrm{pd}}$. Since dipolar mechanisms have shorter relaxation times than space charge mechanisms, samples partially discharged at increasing $T_{\mathrm{pd}}$ temperatures become less polarized by dipolar mechanisms while space charge mechanisms are less affected. This procedure is, therefore, well suited to study the role of dipolar mechanisms in thermal poling. An alternative would have been to employ fractional poling but this would result in a weaker signal because this poling technique yields less polarization.  

Once at room temperature, polarization is relatively stable and additional procedures can be performed on the sample. The sample is further cut in a circular shape to let it fit inside the LIPP sample holder. The diameter of the resulting sample is one millimeter larger than the aluminum electrode. This margin would not be enough to pole it again but is enough to discharge it later so the mechanisms activated in the sample can be revealed.

\begin{figure}
\begin{center}
\includegraphics[width=12cm]{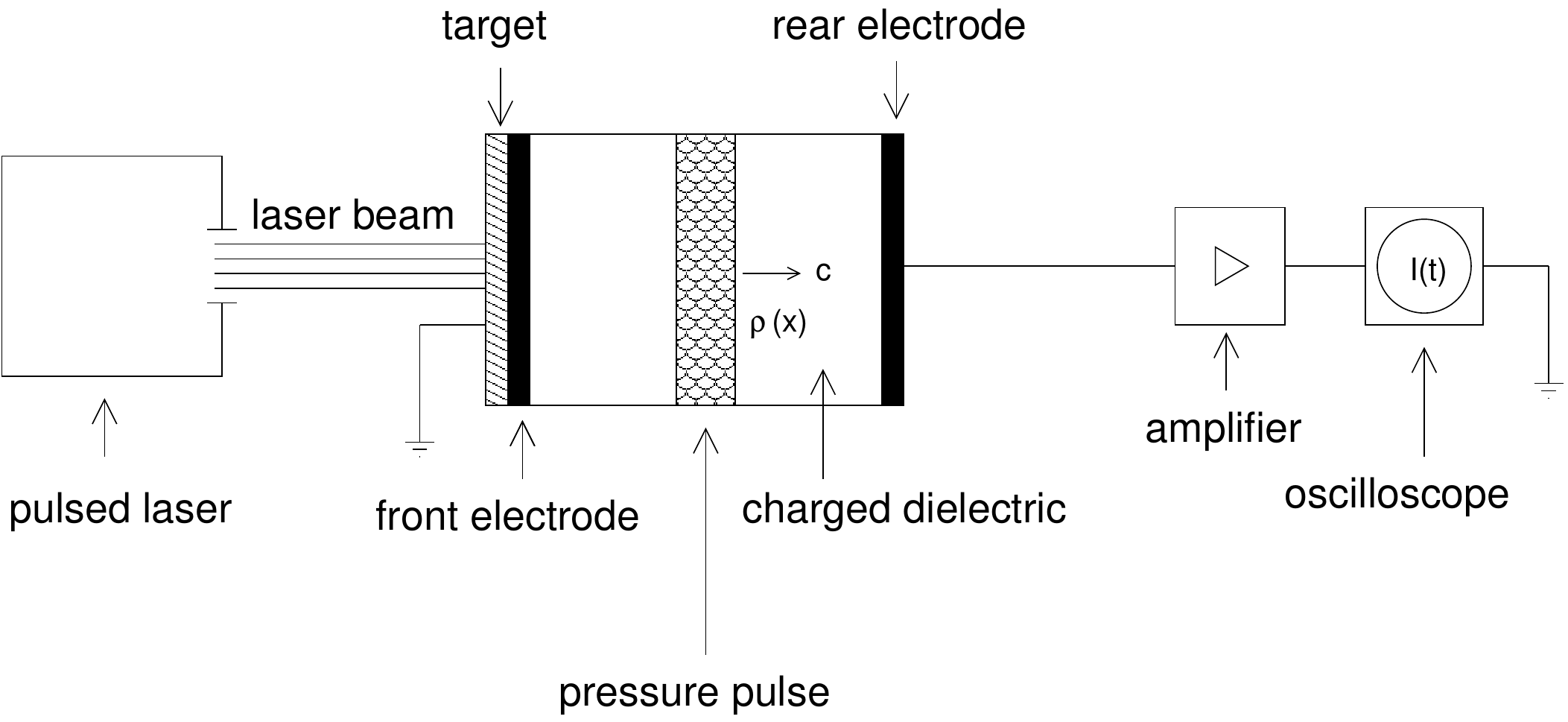}
\caption{Scheme of the LIPP setup}\label{lipp-setup}
\end{center}
\end{figure}

After this preparation step, the charge profile of the samples is measured with the LIPP method. A typical LIPP setup can be seen in Figure~\ref{lipp-setup}. 
Within this method a laser pulse hits a target producing ablation. The momentum of the ablated particles is transferred, because of the momentum conservation law, to the sample as a pressure pulse that propagates at the speed of sound. The compression of the sample produces a change in its capacity. Since the sample is shortcircuited through a current amplifier, the change in capacity leads to a current due to the redistribution between electrodes of their free charge.

Laser pulses with a duration of $8$~ns, wavelength of $1064$~nm and an energy of $10$~mJ were produced by a Nd:YaG pulsed laser. 

% EVA: Etthylen vinyl acetate (EVA ALCUDIA PA 539, Repsol, Spain) with a melt flow index of 1.18 g/min and density of 937 kg/m3.

As a target we have used a blend of Ethylene vinyl acetate (EVA) with carbon black. The components were mixed using a Brabender mixer type W 50 EHT PL (Brabender GmbH \& Co. KG, Germany) heated at $120$~$^\circ$C and a mixed speed of $75$ rpm. The EVA matrix were melted for a minute and then, the black carbon were added and mixed for a period of 10 min. The blend was then consolidated in a hot plates press machine type Collin Mod. P 200E (Dr. Collin GmbH, Germany) forming square sheets, measuring $160 \times 160 \times 0.3$~mm$^3$. Consolidation was carried out at a pressure of $100$~kN for $5$~min using temperatures of 100~$^\circ$C. Finally, the square sheets were cooled under pressure using cool water. The sheets were cut into disks of $1$~cm of diameter and $10$~$\mu$m of thickness. 

To ease the transference of the pressure pulse between the target and the sample a small layer of silicon oil has being put between them. Also, a small drop of silicon oil is placed over the target to increase its lifetime. It has been checked using a piezoelectric that this results in a pressure wave of roughly triangular shape with a base--width of $35$~ns and a half--width of $20$~ns.

The target and the sample are hold together in an especially designed sample holder. The sample holder ground electrode has an opening so the laser can impact the target. Since the target itself is conductive it does not impede electrical contact between the sample holder electrode and the vacuum deposited electrode on the front side of the sample. At the opposite side of the sample holder there is another electrode in direct contact with the vacuum deposited electrode on the rear side of the sample. Throughout experiments, the side of the sample that was exposed to the positive pole when poling is the front side.

Because of the intensity and duration of the current pulse, a high speed current amplifier had to be used. In our setup it was a FEMTO DHPCA high speed variable gain amplifier.

The current that flows through the current amplifier is converted into a voltage that is read by an oscilloscope. This signal is interpreted according to the following equations \cite{migliori80, sessler82}.
\begin{equation}
I(t) = - (2 - \epsilon_r^{-1}) \chi \frac{A}{s} c \int_0^s P(ct - z) \rho(z) dz
\end{equation}
where $\epsilon_r$ is the relative permittivity, $\chi$ is the compressibility of the sample, $A$ is the sample area, $s$ is the thickness of the sample, $c$ is the speed of sound in the sample, $P(z)$ is the wavefunction of the pressure pulse and $\rho(z)$ is the charge profile. If we assume that $P(z)$ is a sharp impulse we obtain
\begin{equation}
I(t) \simeq - (2 - \epsilon_r^{-1}) \chi \frac{A}{s} c^2 \tau P \rho(ct)
\end{equation}
where $P$ is the amplitude of the pulse and $\tau$ is its duration. Since the intensity through the circuit is proportional to the voltage measured by the oscilloscope, this voltage is approximately proportional to the charge profile through the variable change $x = ct$.

\begin{figure}
\begin{center}
\includegraphics[width=12cm]{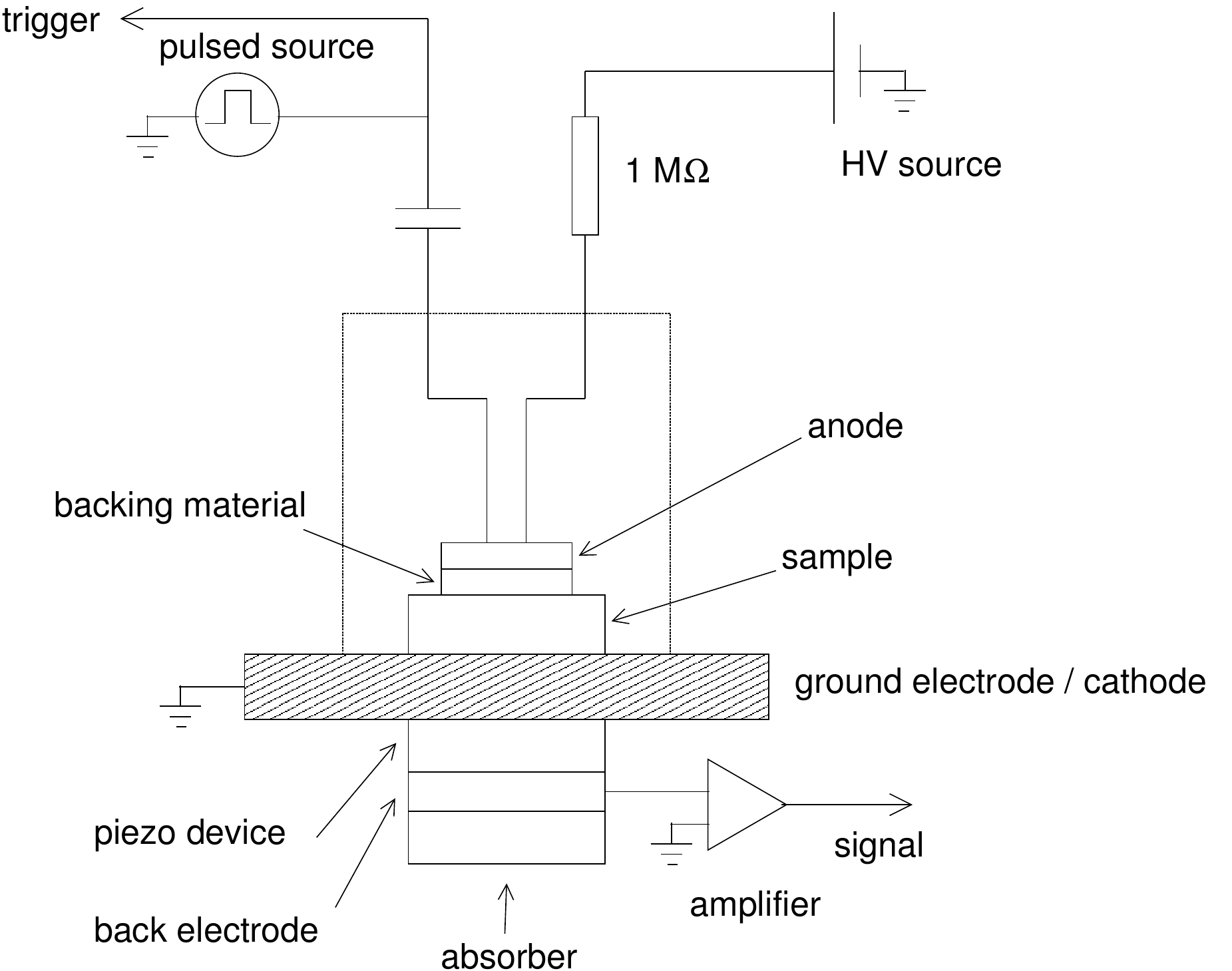}
\caption{Scheme of the PEA setup}\label{pea-setup}
\end{center}
\end{figure}

To allow for an easier interpretation of results and compare the merits of both techniques, the charge profile has also been measured with the PEA method. A commercial setup provided by TechImp (Italy) was employed. A typical PEA setup can be seen in Figure~\ref{pea-setup}. 
Within this method, an electrical pulse (in our setup with a width of $20$~ns and an amplitude of $300$~$V$) is applied between the anode and the cathode. This pulse produces a movement in the charges of the system. The result is an acoustic pulse that is recorded through a piezoelectric. Simultaneously, a direct current high voltage can be applied to the sample to pole it in--place but we have used this possibility only in the measurement of the speed of sound in PET. All the other samples where thermally poled by means of the TSDC setup. 

The recorded signal can be deconvolved in order to improve its resolution but since it is not easy to obtain optimal results and we were seeking just qualitative insight we present the PEA results without deconvolution.

Once the charge profile measurements are performed, the sampled is completely discharged to analyze by TSDC the charge that has been measured by LIPP and PEA.

\begin{figure}
\begin{center}
\includegraphics[width=12cm]{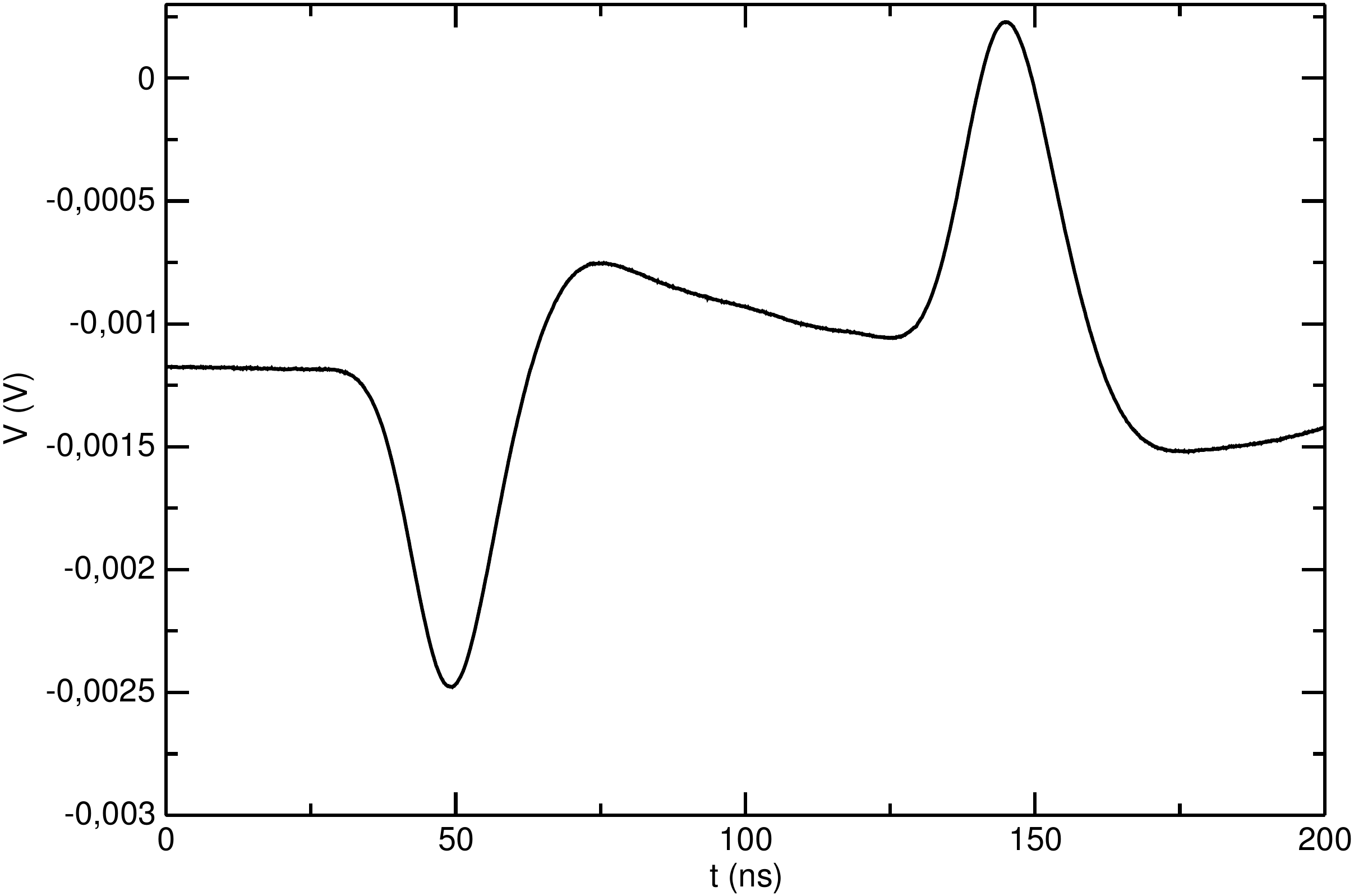}
\caption{PEA measurement of the charge profile of a PET sample while 10~kV are being applied at room temperature}\label{sos}
\end{center}
\end{figure}

Also, because it is easier to locate the limits of the sample with PEA, we have used this technique to measure the speed of sound in our samples. This can be seen in Figure~\ref{sos}.
This figure presents the signal obtained when a $10$~kV field is applied to a sample at room temperature. The position of the mirror charges can be used to locate the limits of the sample and, therefore, to calculate the speed of sound since we know the distance between them. From these measurement, a speed of $2.4$~km/s is deduced.

\section{Results and discussion}

\begin{figure}
\begin{center}
\includegraphics[width=12cm]{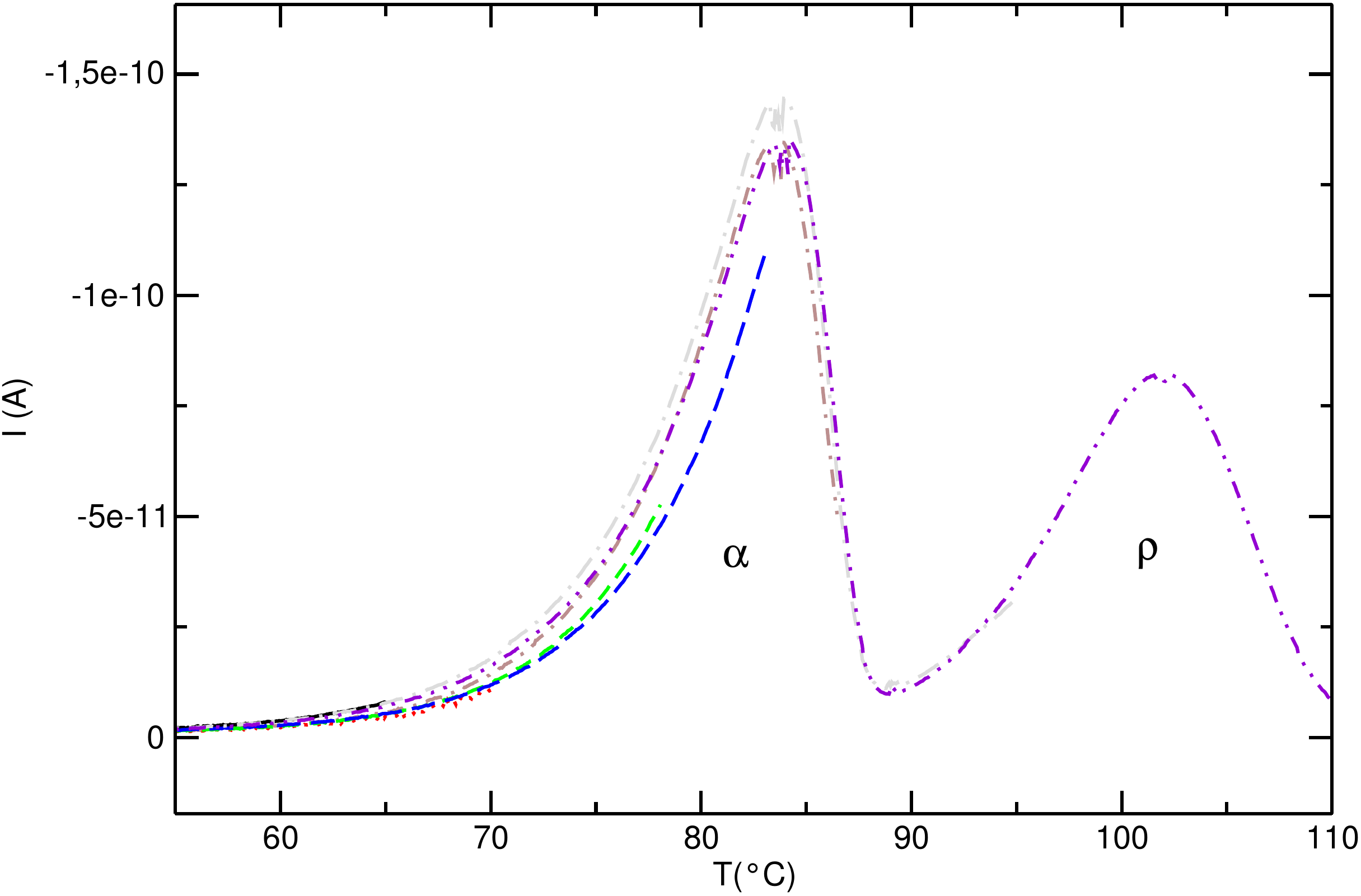}
\caption{Current recorded during partial discharges for the following values of $T_{\mathrm{pd}}$: 65~$^\circ$C (continuous), 70~$^\circ$C (dotted), 78~$^\circ$C (dashed), 83~$^\circ$C (long--dashed), 86~$^\circ$C (dot--dashed), 95~$^\circ$C (dot--long--dashed), 110~$^\circ$C (dot--dot--dashed).}\label{partial}
\end{center}
\end{figure}

As explained in the previous section, the sample is partially discharged after being poled and before its charge profile is measured. The current recorded during the partial discharges is plotted in front of temperature in Figure~\ref{partial}. 
this figure represents the depolarization of each sample during the partial discharge. 

In this figure we can see the $\alpha$ (dipolar) relaxation mostly between $70$ and $87$~$^\circ$C and the $\rho$ (free space charge) relaxation that starts at approximately $88$~$^\circ$C. For the samples with $T_{\mathrm{pd}} = 65$ and $70$~$^\circ$C the intensity plot ends before the $\alpha$ shows up. For this sample we expect both dipolar and free space charge polarization. The plots that correspond to samples with $T_{\mathrm{pd}} = 78$ and $83$~$^\circ$C show a portion of the $\alpha$ peak, that is almost fully discharged for $T_{\mathrm{pd}} = 86$~$^\circ$C. Therefore we expect that these samples have progressively less dipolar polarization while maintaining their free space charge polarization. The sample with $T_{\mathrm{pd}} = 95$~$^\circ$C will have no dipolar polarization, according to this plot, but should retain most of its free space charge. Finally, the sample with $T_{pd} = 110$~$^\circ$C should have almost no polarization of either type.

\begin{figure}
\begin{center}
\includegraphics[width=12cm]{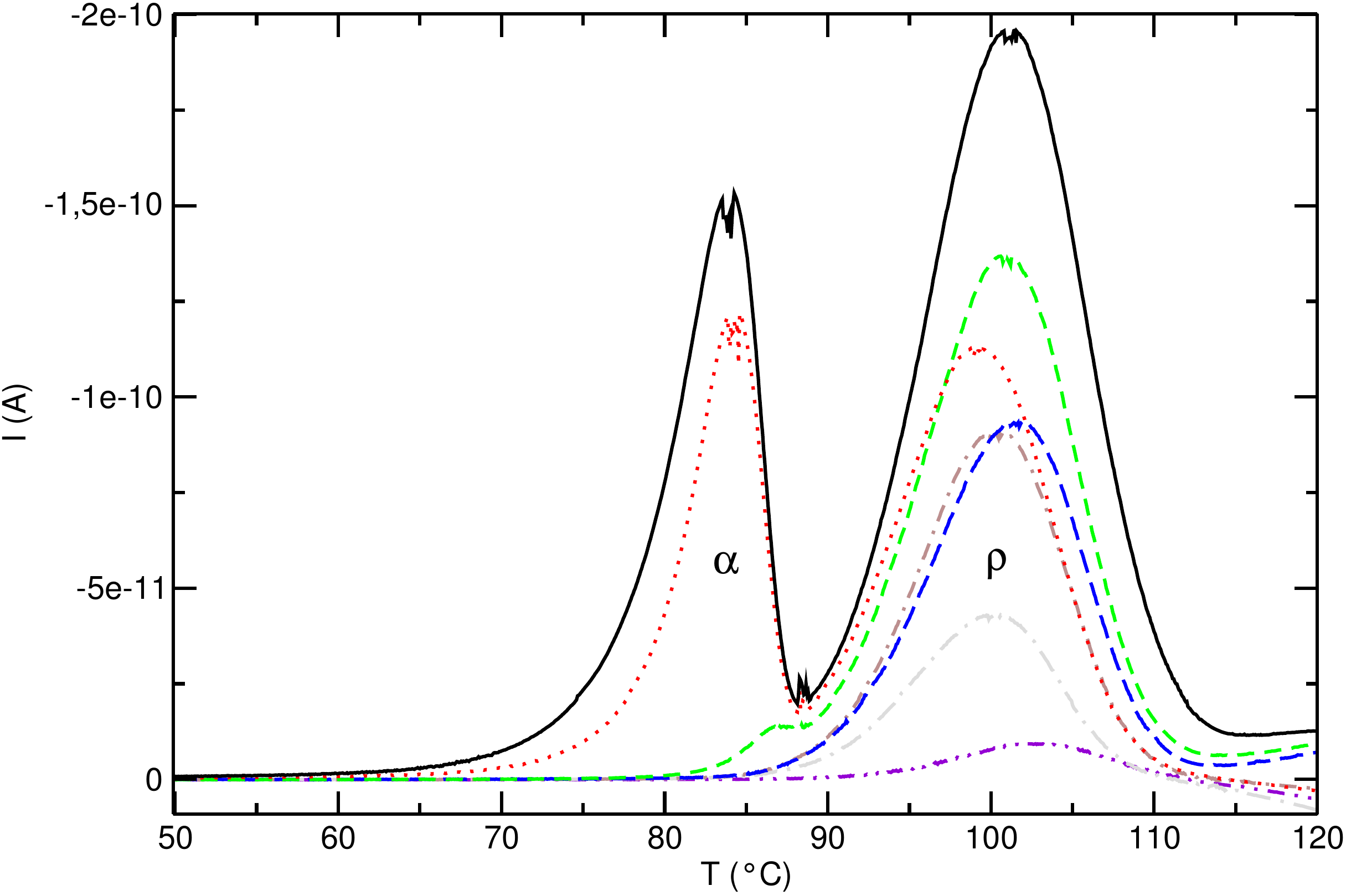}
\caption{Current recorded during final discharge for the following values of $T_{\mathrm{pd}}$: 65~$^\circ$C (continuous), 70~$^\circ$C (dotted), 78~$^\circ$C (dashed), 83~$^\circ$C (long--dashed), 86~$^\circ$C (dot--dashed), 95~$^\circ$C (dot--long--dashed), 110~$^\circ$C (dot--dot--dashed).}\label{full}
\end{center}
\end{figure}

To confirm these assumptions, the samples are also discharged, this time fully, after the charge profile has been measured. In this way we can analyze the actual charge that was present during the charge profile experiments. The intensity plots in front of temperature of these discharges are presented in Figure~\ref{full}.

These plots confirm that samples with $T_{\mathrm{pd}} < 75$~$^\circ$C retain a large portion of dipolar and space charge polarization. The $T_{\mathrm{pd}} = 78$~$^\circ$C shows only a small portion of dipolar polarization. This means that some depolarization occurs when the sample is cooled after it has reached $T_{\mathrm{pd}}$ and the current is no longer recorded. Moreover, the $\rho$ peak is also reduced greatly and a homopolar current appears at a slightly higher temperature. This behavior can be observed for samples with $T_{\mathrm{pd}} = 86$, $89$, and $95$~$^\circ$C. The most probable cause of this effect is charge injection from the electrodes.

\begin{figure}
\begin{center}
\includegraphics[width=12cm]{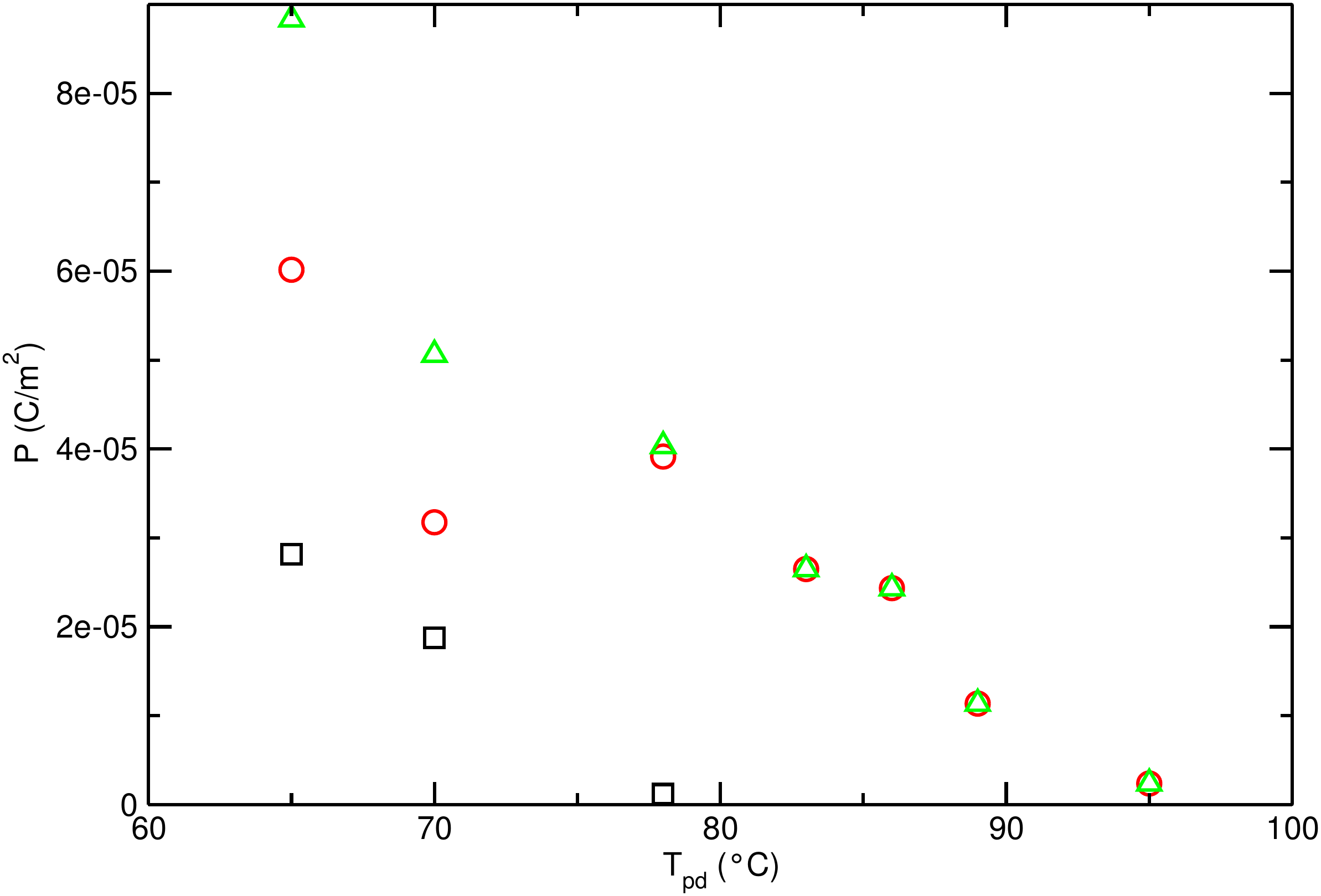}
\caption{Polarization of PET samples for a given $T_{\mathrm{pd}}$: dipolar polarization (square), free space charge polarization (circle), total polarization (triangle).}\label{polarization}
\end{center}
\end{figure}

We can obtain the polarization by mechanism integrating the area of each peak and, in this way, present graphically how $T_{\mathrm{pd}}$ affects the polarization of the sample. This representation is presented in Figure~\ref{polarization}
This is the polarization of the charge whose profile we will be studying. From $78$~$^\circ$C up to higher temperatures we only have $\rho$ polarization while under this temperature we also have $\alpha$ polarization. It can be observed that the space charge polarization points have a higher dispersion which is due to a greater non--repetitivity of space charge experiments.

\begin{figure}
\begin{center}
\includegraphics[width=12cm]{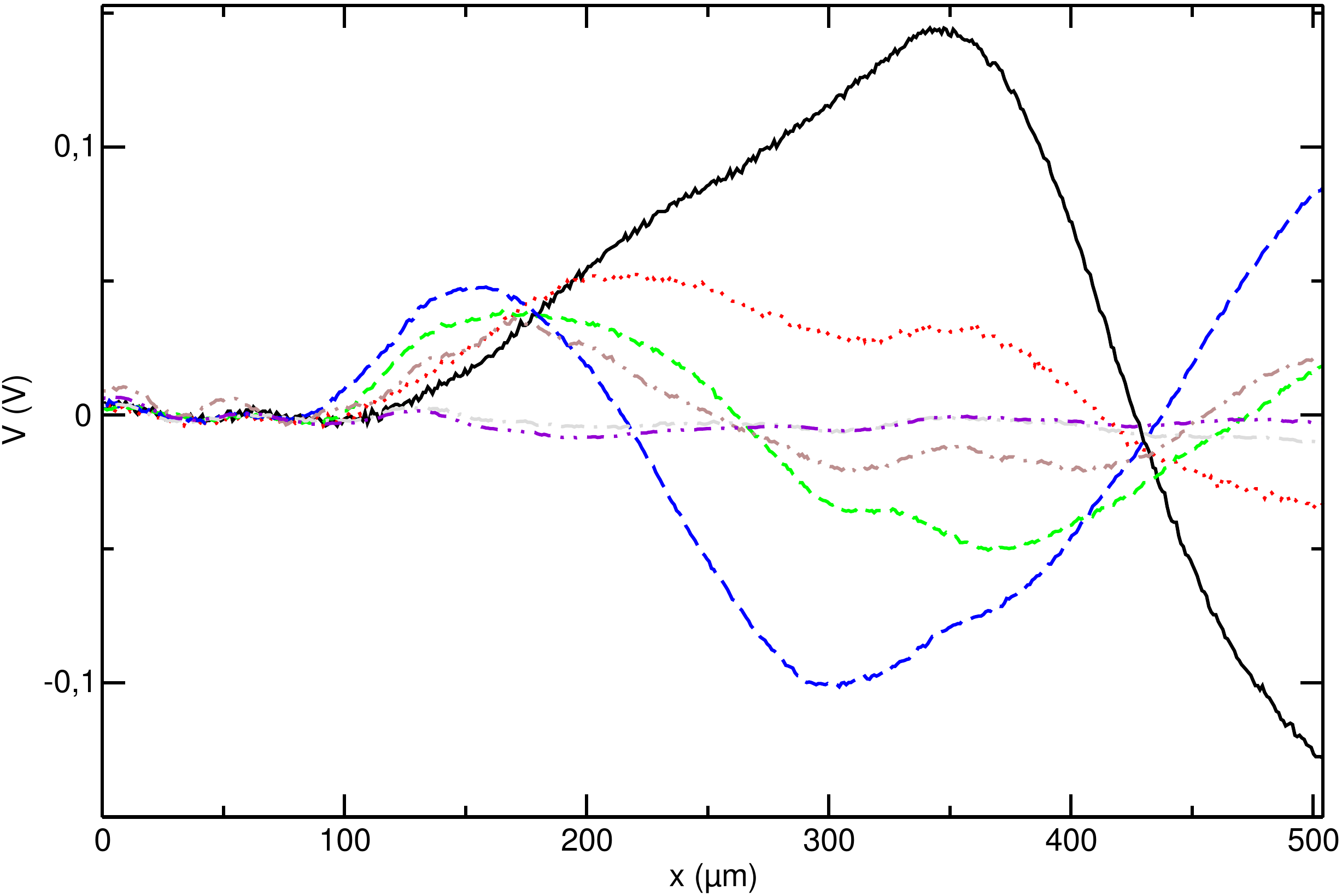}
\caption{Charge profile measured by LIPP for the following values of $T_{\mathrm{pd}}$: 65~$^\circ$C (continuous), 70~$^\circ$C (dotted), 78~$^\circ$C (dashed), 83~$^\circ$C (long--dashed), 86~$^\circ$C (dot--dashed), 95~$^\circ$C (dot--long--dashed), 110~$^\circ$C (dot--dot--dashed).}\label{profile}
\end{center}
\end{figure}

In the LIPP experiments, the current intensity produced when the pressure pulse travels through the sample is converted into a voltage by the current amplifier and represented in Figure~\ref{profile}.
This voltage resembles the charge profile once the time variable has been converted into space using the speed of the pressure pulse. For $T_{\mathrm{pd}} = 65$ and $70$~$^\circ$C the charge profile is asymmetric while for $T_{\mathrm{pd}} = 78$, $83$ and $86$~$^\circ$C it is antisymmetric. Samples with $T_{\mathrm{pd}} = 89$~$^\circ$C and above do not present a meaningful signal. It seems that the charge profiles that correspond to $T_{\mathrm{pd}}$ values that leave only the $\rho$ relaxation activated are antisymmetric. Instead, charge profiles with both the $\alpha$ and the $\rho$ relaxation activated are asymmetric. This is rather surprising since we expect that the signature of dipolar polarization is two narrow peaks of opposite sign at the borders of the sample, which are not seen on the figure, and not a rather broad asymmetric distribution. 

\begin{figure}
\begin{center}
\includegraphics[width=12cm]{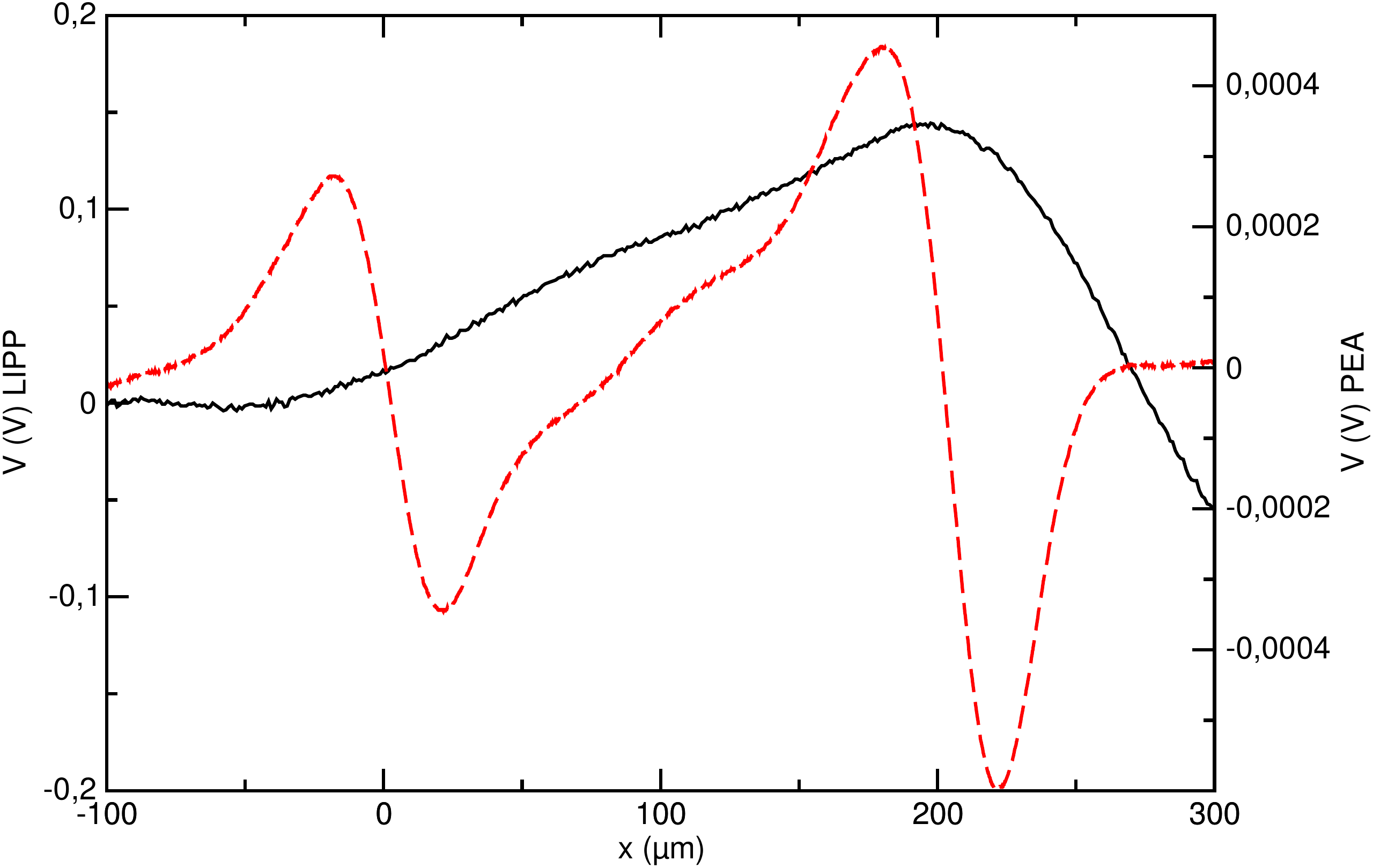}
\caption{Comparison between LIPP (continuous) and PEA (dashed) charge profile for $65$~$^\circ$C}\label{065}
\end{center}
\end{figure}

To interpret further these results we compare LIPP with PEA results. For clarity, we begin comparing two single plots that correspond to the sample ($T_{\mathrm{pd}} = 65$) measured by LIPP and PEA. Figure~\ref{065}
presents this comparison. A good agreement can be observed but with some caveats. The sharp distribution at the edges of the sample can only be seen by PEA, together with its mirror images at the electrodes. It seems clear that our LIPP setup, as it has been observed by other researchers \cite{takada98}, is not able to detect thin charge distributions, especially if they are neutralized by a nearby and similarly thin charge distribution of opposite sign. Since in LIPP the current source is the same sample, the capacity of the sample (and of the sample holder) tends to integrate the signal, smoothing abrupt changes and giving only an average value. PEA does not suffer from this inconvenient because the electrical signal is just the record of the transducer and therefore its resolution is limited only by the duration of the voltage pulse.

\begin{figure}
\begin{center}
\includegraphics[width=12cm]{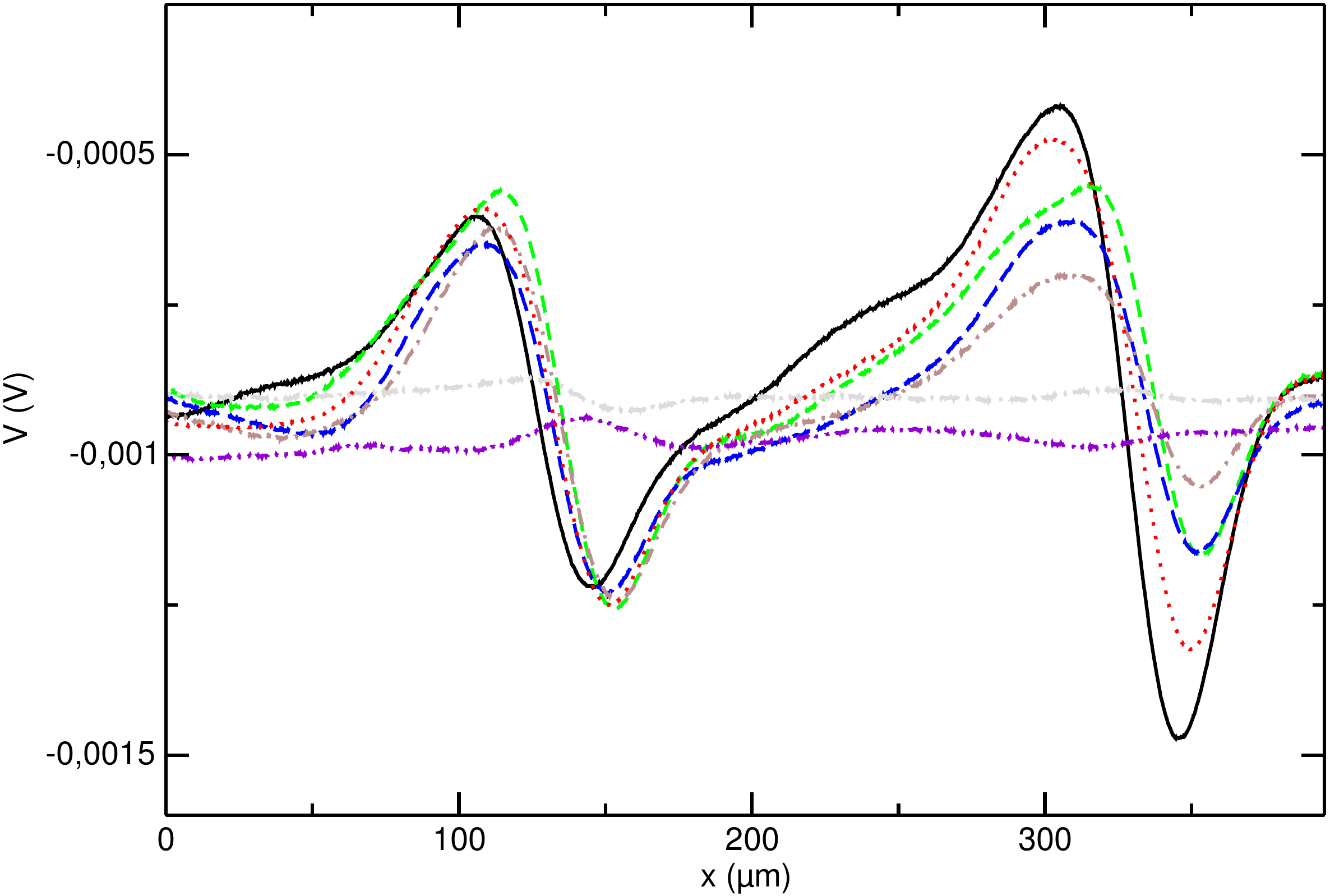}
\caption{Charge profile measured by PEA for the following values of $T_{\mathrm{pd}}$: 65~$^\circ$C (continuous), 70~$^\circ$C (dotted), 78~$^\circ$C (dashed), 83~$^\circ$C (long--dashed), 86~$^\circ$C (dot--dashed), 95~$^\circ$C (dot--long--dashed), 110~$^\circ$C (dot--dot--dashed).}\label{pea}
\end{center}
\end{figure}

The PEA results that can be seen in Figure~\ref{pea}
correspond to the same samples previously measured by LIPP. Measurements were taken before the final full discharge. More or less the same pattern as with LIPP is confirmed (asymmetric profile when $T_{\mathrm{pd}}$ is below $78$~$^\circ$C and antisymmetric profile for this temperature and higher ones) but now we can detect additional features of the charge profile. There is no change in the anode side until the $\rho$ peak is discharged. This means that the charge distribution induced by dipolar polarization is just too narrow to show up even with PEA \cite{hoang14}. The cathode side presents a larger signal intensity that decreases to change the charge profile from asymmetric to antisymmetric. In fact, similar profiles have been reported previously in other materials with other methods \cite{mudarra99}.

It seems clear that the main antisymmetric charge distribution corresponds to the $\rho$ peak but the fact that polarization is not uniform may be due to several reasons. A plausible explanation can be that it is because the displacement of charges is larger than in the case of dipolar orientation and, therefore, the presence of the border has a certain influence in the polarization at each point. 

The evolution of the asymmetry can be due to the following. In PET there is injection of electronic carriers from the cathode \cite{ieda87}. This injection takes place at the cathode even for low electric field values \cite{neagu09} and may be aided by the relatively high poling temperature \cite{zhenlian13}. This injected negative charge may favor the buildup of positive heterocharge near the cathode. This additional heterocharge would be localized in shallow traps and should be released at a lower temperature than ordinary heterocharge. This would explain why charge profile is more intense near the cathode and also why this distribution is the first one to relax.

\section{Conclusions}

The most immediate conclusion about the results is that thermal poling is more complex than expected since it does not only involve dipolar and space charge poling mechanisms but also injected charge from the electrodes. This should be taken into account whenever the reproducibility or stability of a thermally--poled electret is required.

As a consequence the shape and, especially, the evolution of the charge profile of thermally--poled electrets is rather surprising and not easy to explain. We found our results compatible with charge injection of negative carriers at the cathode, as, otherwise, suggested by the literature. Even though there is not enough experimental information to determine the exact charge dynamics in the material, we can make a some assumptions about what happens in the material, mainly that injected negative charge favors the trapping of additional heterocharge in shallow traps. This would give an asymmetrical shape to the charge profile that corresponds to the $\rho$ relaxation with a higher density near the cathode.

In the LIPP results is is not possible to observe image charge at the electrodes, unlike in the PEA results. Since the duration of the pulses are comparable in both setups this must be a consequence of the arrangement of our sample holder. In neither case the observed charge profile is related to dipolar polarization, that seems to be too uniform to be detected. 

PEA is definitively more sensible to the different types of charge mechanisms activated in the sample giving a more complete picture of the charge profile. Nevertheless, this may be a disadvantage when trying to discern between the origin of the charge profile. Instead, LIPP seems to do rather well to detect injected charge. LIPP may be more practical for studying space charge with less interference from other kind of charge mechanisms and in combination with PEA it may be useful to distinguish the origin of the charge. In this sense we may well say that these techniques are complementary.

\bibliographystyle{unsrt}
\bibliography{lipp}

\end{document}